\documentclass[twocolumn,prl]{revtex4}
\usepackage{graphicx}
\usepackage{amsmath,amsthm,bbm,times,mathtools}
\usepackage{amssymb}
\usepackage{bm}
\usepackage{bbold}
\usepackage{color}

\usepackage{pst-plot,ulem}

\newcommand{\ran}{\operatorname{ran}}

\newcommand{\gap}{\operatorname{gap}}

\newcommand{\spa}{\operatorname{span}}
\def\tr{\mathop{\mathrm{tr}}\nolimits}

\newcommand{\ml}[1]{\color{blue} { #1} \black }

\newtheorem{theorem}{Theorem}
\newtheorem{lemma}{Lemma}

\date{October 25, 2024}

\begin{document}
	
	\title{The charge gap is greater than the neutral gap in fractional quantum Hall systems}

	\author{Marius Lemm}
	\affiliation{Department of Mathematics, Eberhard Karls University of T\"ubingen, 72076 T\"ubingen, Germany}
	
	\author{Bruno Nachtergaele} 
	\affiliation{Department of Mathematics
		and Center for Quantum Mathematics and Physics,
		University of California, 
		Davis, CA  95616-8633, USA}

	\author{Simone Warzel}
	\affiliation{Department of Mathematics and Physics, TU M\"{u}nchen, 85747 Garching, Germany}
	\affiliation{Munich Center for Quantum Science and Technology, 80799 M\"unchen, Germany}

	\author{Amanda Young}
	\affiliation{Department of Mathematics, University of Illinois Urbana-Champaign, Urbana, IL 61801, USA}
	
	\begin{abstract}
		Past studies of fractional quantum Hall systems have found that the charge gap dominates the neutral gap for all relevant parameter choices.
		We report a wide-ranging proof that this domination is in fact a universal property of any Hamiltonian that satisfies a few simple structural properties: translation-invariance, charge conservation, dipole conservation, and a fractionally filled ground state. The result applies to both fermions and bosons. Our main tool is a new mathematical scheme, the gap comparison method, which provides a sequence of inequalities that relate the spectral gaps in successive particle number sectors. Our finding sheds new light on dipole conservation's profound effects on many-body physics.
	\end{abstract}
	
	\maketitle

	\paragraph{Introduction.} 
	Symmetries characterize physical systems: 
	their presence determines thermodynamic and topological phases, and imposes constraints on the eigenvalues and eigenstates of Hamiltonians. 
	While charge- and translation symmetry have been considered for decades, the impact of an additional dipolar symmetry has only been theoretically investigated as of late. Even more recently, such systems have become experimentally accessible in cold-atom experiments~\cite{Lake:2022aa,Scherg:2021aa,Zahn:2022aa,Kohlert:2023aa}. 
	Among the interesting phenomena caused by the presence of such a dipole symmetry is the appearance of immobile, low-energy excitations known as fractons, and of many-body scars high up the energy spectrum~\cite{You:2018aa,Pollmann,Pretko:2020aa,Morningstar:2020aa,Bernevig10.19,Moudgalya:2022aa,Gorantla:2022aa}. The change of hydrodynamic laws as well as of the critical dimensions for the absence of phase transitions are other features of such topological dipolar insulators~\cite{Gromov:2020aa,Feldmeier:2020aa,PhysRevLett.132.137102,Lake:2023aa,Kapustin:2022aa,HHY34,ZSPK24}.

	Prototypes of systems that exhibit this additional symmetry are fractional quantum Hall systems (FQHS) \cite{Seidel:2005ab}.   These are characterized by the appearance of a maximal fractional filling at which the system becomes incompressible. The accompanying step in the ground-state energy at increased filling is referred to as the charge gap. This energy gap is to be compared to the neutral gap, which is the energy difference between the first excited and the ground-state energy at maximal filling. In FQHS, the charge gap is observed to be larger than the neutral gap, whose eigenstate is identified with the magneto-roton~\cite{PhysRevLett.54.237,Girvin:1986aa,morf2002excitation,bonderson2011numerical,PhysRevLett.112.046602,Lu:2024bh}. This inequality among the two gaps is shared by the nematic fractional Hall phases~\cite{Fradkin:1999aa,Regnault:2017aa, PhysRevResearch.2.033362,Pu:2024aa}. The characteristic of the nematic phase is that the neutral gap vanishes while the charge gap remains open. 
	
	It is a fundamental question of how robust the observed phenomena of incompressibility and the domination of the charge gap over the neutral gap are.
	In this Letter, we report on a proof that these features are universal among Hamiltonians which satisfy a few simple structural properties:  translation-invariance, charge conservation, dipole conservation, and a fractionally filled ground state. While FQHS give examples of Hamiltonians with these structural properties, our proof does not invoke any physical details of such systems, only the symmetries and ground state properties mentioned above. The main new insight behind the proof is the gap comparison method (Lemma \ref{lemgc}) which provides a sequence of inequalities between the spectral gaps in successive particle number sectors.
	
	
	\paragraph{General setup.} 
	Motivated by FQHS in the torus geometry, we consider spinless fermions or bosons on a ring $  [ 1\ml{:}L ] = \mathbb{Z}/L\mathbb{Z} $,
	and the broad class of many-particle Hamiltonians 
	\begin{align}\label{eq:strucH}
		& H = \sum_{m=m_0}^{m_1} H^{(m)}  \\
		& H^{(m)}= \mkern-10mu \sum_{\substack{ 1\leq j_1\dots j_m\leq L \\ 1\leq k_1 \dots k_m\leq L} } W_{j_1\dots j_m}^{k_1 \dots k_m}
		a^\dagger_{j_1} \dots a^\dagger_{j_m}  a_{k_m} \dots a_{k_1}  \notag
	\end{align}
	where  $ m_0 \leq m_1 $ and the $ m $-body coefficients $  W_{j_1\dots j_m}^{k_1 \dots k_m} \in \mathbb{C} $ are only constrained by the requirement that all $ H^{(m)} $ are self-adjoint and satisfy several assumptions below. By $  (a_j^\dagger) $ and $ (a_j )$  we denote the canonical creation and annihilation operators for each site $j\in[1:L]$.
	
	
	Central to our study are the translation, charge, and dipole symmetries. The translation symmetry is given by the unitary
	\[ 
	T^\dagger a_j T = a_{(j-1) \bmod L } \quad \mbox{for all $ j \in [1:L] $},
	\] 
	and charge and and dipole symmetries are encoded by the unitaries
	\[ 
	U = \exp\left( \frac{2\pi i}{L} N \right)  , \quad V  = \exp\left( \frac{2\pi i}{L} D \right)  
	\]
	where $U$ is generated by particle-number operator $ N  = \sum_{j=1}^L  N_j $ with $ N_j= a_j^\dagger a_j  $, and $V$ is generated by the  
	dipole operator $ D = \sum_{j=1}^L j \ N_j  $.
	The interplay of these three symmetries is captured by the relations
	\begin{equation}\label{eq:rel}
		VT = UTV , \qquad UT = TU , \qquad UV = VU .
	\end{equation}
	This algebra and its generalizations have been studied in~\cite{Gromov:2019aa}.\\
	
	We now state our assumptions on the Hamiltonian $H$.
	
	\begin{description}
		\item[Symmetries:] $ H $ commutes with  $ T , U$, and $V$. 
	\end{description}
	The Hamiltonian (\ref{eq:strucH}) as well as its $ m $-body terms are a direct sums 
	$ H^{(m)} =  \bigoplus_{n\geq 0}  H_n^{(m)} $ of  $ n $-particle Hamiltonians. 
 As is the case with FQHS, we furthermore assume there is a particle number $ n_q \in \mathbb{N} $ corresponding to some fractional filling
	\[ \frac{p}{q} =\frac{n_q}{L}   \]
	with $ 1\leq p<q \in \mathbb{N} $ coprime, and that $H_{n_q}$ satisfies:
	\begin{description}
		\item[Positivity:]  $  H_{n_q} \geq 0 $, and\\
		in case $ m_1 > m_0 $, also $ H_{n_q}^{(m)} \geq 0 $ for all $ m \geq m_0 +1 $, 
	\end{description}
	and that the ground state space is given by $ \ker H_{n_q} $ and has the following structure:
	\begin{description}
		\item[Ground-States:] $\;  \ker H_{n_q} = \spa\left\{ \varphi , T  \varphi ,  \dots , T^{q-1} \varphi \right\} $ for some normalized $ \varphi \in\ker H_{n_q} $ such that
		\begin{itemize}
			\item $ \varphi $ is an eigenvector of $ U $ and $ V $.
			\item $ \varphi $ is $ q $-periodic: \quad $ T^q \varphi = \varphi $. 
		\end{itemize} 
	\end{description}
	Note that the presence of the three symmetries is known to force every eigenstate at fractional filling $ n_q $ to be at least $ q $-fold degenerate~\cite{Seidel:2005ab,Burnell:2024aa}. \\

	These assumptions encompass all $ m $-body pseudopotential Hamiltonians in fractional Hall physics. We recall that in the $ 2 $-body case $ m_0=m_1 = 2 $, these take the form~\cite{Trugman:1985lv,Pokrovsky_1985,Duncan1990,Lee:2015aa}:
	\begin{align}\label{eq:HPseudo}
		H  & =  \sum_{s \in \frac{1}{2} [1,2L] } Q_s^\dagger Q_s\\
		Q_s &  = \sum_k^\prime F_s(k)\ a_{(s-k) \bmod L}\  a_{(s+k) \bmod L}  . \notag 
	\end{align}
	Here and in the following, additions are understood modulo $ L $. The $ s $-sum runs over all half- and integers in $ [1,L] $, and the primed $ k $-sum is over half- or integers in $ [0,L] $ depending on the value of $ s $. Hamiltonians of the form~\eqref{eq:HPseudo} are parent Hamiltonians for all basic filling fractions of the form $ 1/q$ (with $ q \geq 3$ odd for fermions, and $ q \geq 2 $ even for bosons, cf.~\cite{Regnault:2003aa}). In this case, $ \varphi $ plays the role of the Laughlin wavefunction~\cite{PhysRevLett.50.1395,Haldane:1985aa}.  The class of Hamiltonians \eqref{eq:strucH} also encompasses all other $ 2$-body pseudopotential Hamiltonians, including those whose ground states go beyond the basic filling fractions, as demonstrated e.g.\ by the Jain-states~\cite{Chen:2017aa,Bandyopadhyay:2020aa}. The general $m$-body form of \eqref{eq:strucH} is relevant in the description of more exotic fractional quantum Hall states such as the Gaffnians which require  higher $ m $-body parent Hamiltonians~\cite{Lee:2015aa}. Finally, truncated pseudopotentials \cite{Bergholtz:2005pl,Nakamura:2012bu,PhysRevB.85.155116,NWY2020b,nachtergaele:2021a,warzel:2021,warzel:2022} are also within our framework, as well as their  limits~\cite{kapustin:2020}. We remark that in all these examples, the Hamiltonian is in fact positive with $ H_n^{(m)} \geq 0 $ for all $ n ,m $. \\
	
	\paragraph{Incompressibility.} 
	In the above setup, the translates  $ T^j \varphi  $,  $ j=0,\ldots,q-1$, form an orthonormal basis of the ground state at $ n_q $ such that
	\begin{equation}\label{eq:gsp}
		\dim  \ker H_{n_q} = q ,
	\end{equation}
	cf.~\cite{Seidel:2005ab,Burnell:2024aa}. 
	This is easily seen by observing that  these states are $ V $-eigenstates with distinct eigenvalues 
	\begin{equation}\label{eq:eigenvaluesV}
		V T^j \varphi  =   \exp\left( 2\pi i  \ \Big( \frac{d_\varphi }{L} + \frac{j}{p} \Big) \right)  T^j \varphi ,
	\end{equation} 
	with $ d_\varphi \in \mathbb{Z} $ the $ D $-eigenvalue corresponding to $ \varphi $.\\

	It is a hallmark of a system's incompressibility that upon adding an extra particle, the ground-state energy rises. Remarkably, all systems in the above setup are incompressible at $n_q$. Indeed, for any $ n > n_q $, the ground-state energy of $H_n$ is above zero, the ground-state energy of $H_{n_q}$. Equivalently: 
	\begin{theorem}[Incompressibility at $ n_q $] \label{thm:nokern}
		For any $ n \geq n_q $ there are no zero energy eigenstates of $ H_{n+1} \geq 0$, i.e. 
		\begin{equation}\label{eq:nogs}
			\dim  \ker H_{n+1} = 0 , 
		\end{equation}
		as long as the system is sufficiently large: $ L \geq q^2/p $ in the case of fermions, and $ L \geq (1+p) q^2/p $ in the case of bosons. 
	\end{theorem}

	\paragraph{Charge  vs.\ neutral gap.}  
	Comparing~\eqref{eq:gsp} and \eqref{eq:nogs}, the ground-state energy at $ n_q $ vs.\ $ n_q+1 $ exhibits a jump. In the case of FQHS, this energy jump, which is referred to as the charge gap, agrees with $ \gap H_{n_q+1} $ where
	\begin{equation}\label{def:gap}
		\gap H_n = \min_{\substack{\psi \perp \ker H_n \\ \, \| \psi \| = 1    }} \langle \psi , H_n \psi \rangle \quad \forall n.
	\end{equation}
	For FQHS, $ \gap H_{n_q} $ is then the neutral gap, which is conjectured to be the lowest excited energy in the system.
	The notation is chosen such that \eqref{def:gap} represents the spectral gap (i.e., the neutral gap) for $ n =n_q $ and  the positive ground-state energy (i.e., the charge gap) for any $ n > n_q $. As a result, we prove:
	\begin{theorem}[Charge vs. neutral gap] \label{thm:main}
		In the case of $q$-commensurate maximal filling, i.e. $ L = \ell q^2 $ for some $ \ell\geq 3  $, the charge gap dominates the neutral gap, for any $ n \geq n_q $
		\begin{align*}
			&\gap H_{n+1} \geq \frac{n_q}{n_q+1-m_0}  \ \gap H_{n_q}  & \mbox{(fermions)} \\
			&\gap H_{n+1} \geq \frac{n_q-p }{n_q+1-m_0}  \  \gap H_{n_q}     & \mbox{(bosons)} 
		\end{align*} 
	\end{theorem} 
	
	A few remarks are in order. (i) For bosons, the bound implies that the charge gap dominates the neutral gap only for $m_0\geq p+1 $.  (ii) For fermions, the bound is conclusive even for $ m_0 = 1 $, which allows for the presence of a suitably adjusted, negative chemical potential $ \mu N  $ in the Hamiltonian.
	(iii) The commensurability condition $ L = \ell q^2 $ can be dropped for $p=2$. 
	
	There is evidence that in geometries without translation invariance finite-size effects may cause a violation of the inequality \cite{balram2024fractional},   but it is expected that the gap domination result always holds in the bulk. Indeed, by considering a sequence of system sizes tending to infinity, one sees that our charge vs.\ neutral gap result extends to the thermodynamic limit.
	

	The onset of the fractional nematic phase is characterized by the vanishing of the neutral gap in the thermodynamic limit, while the charge gap remains open~\cite{Pu:2024aa}. 
	Since such systems are covered by the above theorem, one cannot expect that the neutral gap is generally uniformly positive. 
	For pseudopotentials~\eqref{eq:HPseudo} corresponding to $ 1/q$-filling with small $ q \in \{ 2, 3 , \dots \} $,  it is conjectured that the excitation gap above the ground-state energy is uniformly positive, which is known as ``Haldane's FQHS conjecture''.  Note that a gap bound $ \gap H_{n_q} \geq c>0$ at fractional filling is not a contradiction to Lieb-Schultz-Mattis type results for systems with dipolar symmetry~\cite{Oshikawa:2000aa,Burnell:2024aa}.
	While the conjecture remains open for the pseudopotentials~\eqref{eq:HPseudo}, it was proven in the case of truncated pseudopotentials~\cite{nachtergaele:2021a,warzel:2021,warzel:2022}. For larger values of $ q $, the neutral gap is conjectured to close~\cite{PhysRevLett.54.237}.

	For Haldane pseudopotentials~\eqref{eq:HPseudo}, our lower bounds may be combined and compared to the upper bounds on the charge gap derived in~\cite{Weerasinghe:2016aa}:
	\[
	\gap H_{n_q+1}  \leq \begin{cases} \frac{4p}{q-p}\ \Delta & \hfill \mbox{(fermions)} \\
		\frac{4p}{q+p} \ \Delta & \hfill  \mbox{(bosons)}
	\end{cases}
	\]
	where $ \Delta =  \frac{1}{4}   \sum_{s} \sum_{k }^\prime \left| F^{(m)}(k) \right|^2  $.\\ %
	
	
	\paragraph{Proof of Theorem \ref{thm:nokern}.}
	For each $ m $-body Hamiltonian 
	\begin{equation}\label{IRid}
		H_{n+1}^{(m)} =   \frac{1}{n+1-m}  \sum_{j=1 }^L a_j^\dagger\;  H_{n}^{(m)}\;  a_j 
	\end{equation}
	for all  $ n \geq m $. This follows from the `combinatorial' identity
	\begin{align*}
		N  a^\dagger_{j_1}\cdots a^\dagger_{j_m} a_{k_m}\cdots a_{k_1} & = m a^\dagger_{j_1}\cdots   a^\dagger_{j_m} a_{k_m} \cdots  a_{k_1} \\
		& \hspace{-10pt}+ \sum_{j=1}^L a_j^\dagger \! \left( a^\dagger_{j_1}\cdots a^\dagger_{j_m} a_{k_m}\cdots a_{k_1} \right)\! a_j  
	\end{align*}
	for any $ m \in \mathbb{N} $ and any $ j_1 , \dots , j_m , k_m , \dots , k_1 \in [1:L] $, see also~\cite{Schossler:2022aa}. 
	Applying~\eqref{IRid} separately to each of the $ m $-body terms with $ m \geq m_0 $ and using $ H_n^{(m)} \geq 0 $ for $ m \geq m_0 +1 $, we arrive at the  inductive relation
	\begin{equation}\label{IR}
		H_{n+1} \geq    \frac{1}{n+1-m_0}  \sum_{j=1 }^L a_j^\dagger\;  H_{n}\;  a_j 
	\end{equation}
	valid for all  $ n \geq m_1 $. Note that~\eqref{IRid} implies that $ H_{n}^{(m)} \geq 0 $ for all $ n \geq n_q $ and $ m \geq m_0+1 $. Hence \eqref{IR} holds for all $ n \geq n_q $. 
	As already noted in~\cite{Weerasinghe:2016aa}, one consequence of~\eqref{IR} is that the ground state energy is monotone increasing in $ n $ as long as $ H_n \geq 0 $.
	We remark in passing that, if $H_n^{(m)}\geq 0$ for any $n, m $ (which is satisfied for Haldane pseudopotentials, e.g., \eqref{eq:HPseudo}), then \eqref{IR} shows that $ \dim \ker H_n \geq q $  for any $m_1 \leq  n \leq n_q $, i.e., $ n_q $ is indeed the maximal filling. To see this, observe that \eqref{IR} implies $ \langle a_j \psi , H_{n} a_j \psi \rangle = 0 $  for any $ \psi \in  \ker H_{n+1} $ and all $ j $. The $q$-fold degeneracy then follows from the $V$-symmetry. 
	\\

	We now prove \eqref{eq:nogs}. Given  \eqref{IR} for any $ n \geq n_q $, it suffices to consider the case $ n = n_q $.
	By way of contradiction, suppose that $ \psi \in \ker H_{n_{q}+1} $ is a normalized eigenfunction of $ V $ with eigenvalue $ \exp\left( 2\pi i  \ \frac{d_\psi}{L}  \right) $ for some $ d_\psi \in \mathbb{Z} $. Then, by~\eqref{IR}, we also have $ a_k \psi  \in \ker H_{n_q} $ for all $ k $, and 
	\begin{equation}\label{eq:Vshifta}
		V a_k\psi = \exp\left( 2\pi i  \ \frac{d_\psi - k }{L}  \right)  a_{k} \psi .
	\end{equation}
	This implies that any non-trivial $ a_k \psi \neq 0 $ agrees --- up to a phase and the normalization factor $ \| a_k \psi \| $ --- with one of the vectors in $\left\{ \varphi , T  \varphi ,  \dots , T^{q-1} \varphi \right\} $, and the eigenvalue
	$
	\exp\left( 2\pi i  \ \frac{d_\psi - k }{L}  \right) 
	$
	must coincide with one of the eigenvalues \eqref{eq:eigenvaluesV} of the eigenvectors of $V$ in $\ker H_{n_q}$. There are only $q$ values of $k$ for which this is possible. Hence, $a_k\psi\neq 0$  for at most $q$ values of $k$. Moreover,
	\begin{equation}\label{particlenumberbound}
		n_q+1 = \langle\psi , N \psi\rangle =\sum_{k=1}^L  \langle a_k\psi , a_k\psi\rangle \leq q \max_{k}  \langle\psi , N_k \psi\rangle.
	\end{equation}
	For fermions this implies $n_q+1 \leq q$, a contradiction since $L\geq \frac{q^2}{p}$. 
	
	For bosons, we estimate $ \langle\psi , N_k \psi\rangle$ as follows. First, the $ q $-periodicity of $ \varphi $ implies 
	\begin{equation}\label{eq:boundNj}
		\max_{1\leq j\leq L}  \langle \varphi , N_j  \varphi \rangle \leq  \sum_{j=1}^{q} \langle \varphi , N_j\varphi \rangle = \frac{q}{L}   \langle \varphi , N  \varphi \rangle = p.
	\end{equation}
	Since $a_k\psi \in \ker H_{n_q}$, we thus conclude
	\begin{align*}
		p\geq & \  \frac{\langle a_k\psi, N_k a_k\psi\rangle}{\langle a_k\psi, a_k\psi\rangle}
		=  \frac{\langle \psi, N_k^2 \psi\rangle-\langle \psi, N_k\psi\rangle}{\langle \psi, N_k\psi\rangle} \\
		\geq & \ \frac{\langle \psi, N_k \psi\rangle^2-\langle \psi, N_k\psi\rangle}{\langle \psi, N_k\psi\rangle}
		= \langle \psi, N_k\psi\rangle -1 .
	\end{align*}
	This implies
	$
	\langle \psi, N_k\psi\rangle\leq p+1  $, and so \eqref{particlenumberbound} yields $n_q+1 \leq (p+1)q$, a contradiction to the assumption $n_q \geq(p+1) q $. This concludes the proof of Theorem~\ref{thm:nokern}.\\

	\paragraph{Gap comparison method.}
	The proof of Theorem~\ref{thm:main} is our main contribution and it relies on the following new mathematical insight.
	Let $P_n$ be the orthogonal projection onto $ \ker H_{n} $ and set $P_n^\perp=\mathbbm{1}-P_n$. 
	
	\begin{lemma}[Gap comparison method]\label{lemgc}
		For any Hamiltonian satisfying \eqref{IR} and $ H_n \geq 0 $ with $ n \geq m_1 $:
		\begin{equation}\label{eq:gapin}
			\gap H_{n+1} \geq \frac{ n+1 -  \Big\| P_{n+1}^\perp \sum_{j} a_j^\dagger P_{n} a_j P_{n+1}^\perp \Big\| }{n+1-m_0}\gap H_{n} .
		\end{equation}
	\end{lemma}

	To prove this lemma, we fix any normalized $ \psi \in \ran P_{n+1}^\perp $. Using first the inductive relation \eqref{IR} and then the estimate $ H_{n} \geq  \gap H_{n} \left( 1 - P_{n} \right) $ gives
	
	\begin{align*}
		\langle \psi, H_{n+1} \psi \rangle  & \geq     \frac{\sum_{j =1}^L  \langle \psi, a_j^\dagger (1-P_n) a_j  \psi \rangle }{n+1-m_0}  \gap H_{n}  \\
		& =    \frac{n+1  -  \langle \psi, \sum_{j} a_j^\dagger P_{n} a_j  \psi\rangle }{n+1-m_0} \gap H_{n}
	\end{align*}
	Then, \eqref{eq:gapin} follows from $\psi=P_{n+1}^\perp\psi$ and $\langle \psi,A\psi\rangle\leq \|A\|$.\\

	\paragraph{Proof of Theorem~\ref{thm:main}.}
	We begin from Lemma \ref{lemgc}. Note that $ P_{n} = 0 $ for all $ n \geq n_q $ by~\eqref{eq:nogs}. This already implies $ \gap H_{n+1} \geq \frac{n+1}{n+1-m_0}  \gap H_{n} $ for all $n\geq n_q+1$. Therefore, it suffices to prove the claim for $n=n_q$. 
	In this case, $P_{n_q+1}^\perp= \mathbbm{1}$. Note that $\Big\| \sum_{j=1}^L a_j^\dagger P_{n_q} a_j  \Big\|=\| G \|$, where $G$ is the many-body Gram matrix
	\[
	G_{jk,j'k'} =   \langle  a_j^\dagger T^{k-1} \varphi,  a_{j'}^\dagger T^{k'-1} \varphi\rangle,
	\]
	indexed by $ j ,j'\in[1:L]$, $ k,k' \in [1:q] $. 
	
	To bound $\|G\|$,  we again use symmetry considerations. Denoting the $D$-eigenvalue of $ \varphi $ by $ d_\varphi  $, we have 
	\[ V  a_j^\dagger T^{k-1} \varphi =  \exp\left( 2\pi i \frac{d_\varphi + (k-1) n_q +j }{L}\right) a_j^\dagger T^{k-1} \varphi  .
	\]
	This implies that 
	$ \langle  a_j^\dagger T^{k-1} \varphi,  a_{j'}^\dagger T^{k'-1} \varphi\rangle = 0  $ unless $ \left( k n_q +j \right) \bmod L = \left( k' n_q +j' \right) \bmod L  $. 
	The Gram matrix is hence block diagonal  
	\[ 
	G =  \bigoplus_{\gamma = 1}^L \ G(\gamma)  
	\] 
	with $ q\times q $ blocks  
	\begin{multline*}
		G(\gamma)_{k,k'}   =\\  \left\langle a_{(\gamma-n_q k )\bmod L}^\dagger \ T^{k -1} \varphi ,  a_{(\gamma-n_q k' )\bmod L}^\dagger \ T^{ k'-1} \varphi \right\rangle . 
	\end{multline*}
	In turn,  the (anti-)commutation rules imply that each block has the following structure $
	G(\gamma)   =  \mathbbm{1} \pm F(\gamma) $ 
	with $ (-) $ for fermions and  $ (+) $ for bosons and, by translating,
	\begin{align*}
		F(\gamma)_{k,k'} =    \langle &a_{(\gamma+n_qk )\! \!\bmod\! L} \ T^{k -1 +(k+k') n_q} \varphi ,\\  &a_{(\gamma+n_q k') \!\!\bmod \! L} \ T^{ k'-1+ (k+k') n_q} \varphi \rangle.
	\end{align*}
	Note that $n_q=p L/q=p\ell q$ is a multiple of $q$. Hence, the $ q $-periodicity of $ \varphi $ implies that 
	$ F(\gamma) $ 
	is another Gram matrix, such that $ F(\gamma) \geq 0 $. For fermions, this already suffices because it implies $ 0 \leq    G(\gamma)   \leq    \mathbbm{1}  $ and so $\|G\|\leq 1$.
	
	For bosons, \eqref{eq:boundNj} gives
	\begin{align*} \tr F(\gamma) = & \sum_{k=1}^q\left\langle \varphi , N_{(\gamma+1 -(n_q+1) k) \bmod p} \ \varphi \right\rangle \\
		= &  \sum_{k=1}^q\left\langle \varphi , N_{k} \ \varphi \right\rangle = p.
	\end{align*}
	This implies $ 0 \leq F(\gamma) \leq p \mathbbm{1} $ and hence $ 0 \leq    G(\gamma)   \leq  (1+p)  \ \mathbbm{1}  $, which proves Theorem~\ref{thm:main}. \\
	
	
	\paragraph{Conclusion.}
	In this Letter, we presented a  proof that the domination of the neutral gap by the charge gap is in fact a universal phenomenon that is a consequence of a few structural properties: translation-invariance, conservation of charge and dipole moment, and fractional filling of the ground state. It applies to both fermions and bosons. Examples of systems satisfying these properties are general $m$-body pseudopotential models for FQHS. A consequence of our result is that any lower bound on the neutral (Haldane) gap in a FQHS immediately implies a strong, quantitative form of incompressibility, with the same numerical lower bound for the gap.
	
	The main new insight that we exploit is the gap comparison method captured by
	Lemma \ref{lemgc}, which provides an inequality relating the gaps in successive particle number sectors. Looking beyond this paper, we believe that the gap comparison method is a new, simple, and analytically exact tool to study a broad array of spectral gap problems in FQHS. 
	Indeed, as long as $ H_n \geq 0 $, \eqref{eq:gapin} allows to iteratively lower bound the gap starting from its value for small (e.g. $ n= m_1 $) particle numbers. In particular, this suggests a fresh approach to the famous Haldane FQHS gap conjecture as for any $n\leq n_q$,
	\[
	\Big\|P^\perp_{n+1} \sum_{j=1}^L a_j^\dagger P_{n} a_j  P^\perp_{n+1}\Big\|=\| G^{(n)} \|,
	\]
	where $G^{(n)}$ is the $q_n L\times q_n L$ many-body Gram matrix with entries
$$
G^{(n)}_{\alpha j ,\beta k} := \langle a_j^\dagger  \varphi_\alpha  , (1-P_{n+1}) \ a_k^\dagger  \varphi_\beta \rangle
$$
and $ \left( \varphi_\alpha \right)_{1\leq \alpha\leq q_n} $ denotes an orthonormal basis of $ \ker H_{n} $. If a bound $\| G^{(n)} \| \leq 2 + s_{n-1} $ is proved with $ \sum_{3\leq n\leq n_q} s_n/ n$ bounded independently of $n_q$, then a positive lower bound for the Haldane gap follows by iterating \eqref{eq:gapin}. This provides a very concrete approach to the Haldane gap problem. A similar induction on particle-number strategy was successfully employed for a different problem in \cite{Carlen:2003aa}.
	
	
	
	\paragraph{Acknowledgments.}
	This work was supported by the  DFG under the grants TRR 352 – Project-ID 470903074 (ML, SW) and EXC-2111-390814868 (SW), and by the National Science Foundation under grant DMS-2108390 (BN). All authors acknowledge support through a SQuaRE collaboration grant from the American Institute of Mathematics at Caltech, where part of this work was accomplished. BN and AY thank the TU Munich for kind hospitality during a visit in the course of this work. We thank Ting-Chun Lin for useful comments. 
	\bibliographystyle{abbrv}  
	\bibliography{FQHE} 
	
\end{document}